\documentclass[%
 aip,
 jcp,%
 amsmath,amssymb,
 reprint,%
 noeprint,
]{revtex4-1}

\usepackage{mhchem}
\usepackage{graphicx}
\usepackage{dcolumn}
\usepackage{bm}
\usepackage{hyperref}
\usepackage{soul,color}
\newcommand{\grun}{Gr\"{u}neisen }
\newcommand{\mob}{cm$^2$V$^{-1}$s$^{-1}$}

\begin{document}


\title[]{Perspective: Theory and simulation of hybrid halide perovskites}

\author{Lucy D. Whalley}
 \affiliation{Department of Materials, Imperial College London, Exhibition Road, London SW7 2AZ, UK}

\author{Jarvist M. Frost}
 \affiliation{Department of Materials, Imperial College London, Exhibition Road, London SW7 2AZ, UK}
 \affiliation{Department of Chemistry, University of Bath, Claverton Down, Bath BA2 7AY, UK}

\author{Young-Kwang Jung}
 \affiliation{Global E$^3$ Institute and Department of Materials Science and Engineering, Yonsei University, Seoul 03722, Republic of Korea}

\author{Aron Walsh}
 \email{a.walsh@imperial.ac.uk}
 \affiliation{Department of Materials, Imperial College London, Exhibition Road, London SW7 2AZ, UK}
 \affiliation{Global E$^3$ Institute and Department of Materials Science and Engineering, Yonsei University, Seoul 03722, Republic of Korea}

\date{\today}

\begin{abstract}
Organic-inorganic halide perovskites present a number of challenges for first-principles atomistic materials modelling. Such `plastic crystals' feature dynamic processes across multiple length and time scales. These include:
(i) transport of slow ions and fast electrons;
(ii) highly anharmonic lattice dynamics with short phonon lifetimes;
(iii) local symmetry breaking of the average crystallographic space group;
(iv) strong relativistic (spin-orbit coupling) effects on the electronic band structure;
(v) thermodynamic metastability and rapid chemical breakdown.
These issues, which affect the operation of solar cells, are outlined in this perspective. 
We also discuss general guidelines for performing quantitative and predictive simulations
of these materials,
which are relevant to metal-organic frameworks and other hybrid semiconducting, dielectric and ferroelectric compounds.
\end{abstract}

\maketitle

The perovskite mineral, \ce{CaTiO3}, is the archetype for the structure of many functional materials.\cite{Schaak2002}
Metal halide perovskites have been studied for their semiconducting properties since the 1950s\cite{Moller1958}. 
Only recently have organic-inorganic perovskites such as \ce{CH3NH3PbI3} (MAPI) been applied to solar energy conversion, showing remarkably strong photovoltaic action for a solution processed material.\cite{Kojima2009}
The field has progressed rapidly in the last five years. The increase in power conversion efficiency is supported by over three thousand research publications.\cite{Stranks2015b,Saparov2016b,Park2016,Walsh2016,Wallace2017}
Other potential application areas of these materials include 
thermoelectrics,\cite{He2014,Mettan2015}
light-emitting diodes,\cite{Protesescu2015,Stranks2015b}
and
solid-state memory.\cite{Yoo2015a,Liu2017}

Recently we published a short review on the nature of chemical bonding in the these materials,\cite{Walsh2015}
and on the multiple timescales of motion.\cite{Frost2016a}
We will not repeat that material here. 
There is also a recent review from Mattoni and co-workers focusing upon the use of molecular dynamics simulations.\cite{Mattoni2017}

In this Perspective, we address recent progress and current challenges in theory and simulation of hybrid halide perovskites. 
We pay particular attention to predicting properties that assess the photovoltaic potential of a material. 
Factors to consider include: light absorption, charge transport, absolute band energies, defect physics and chemical stability. 
The total energy, electronic energy levels, dielectric function and band effective masses can be calculated with electronic structure methods on a representative (static) crystal structure. 
Lattice and molecular dynamics can describe a variety of dynamic behaviour at finite temperature. 
These perovskites combine a complex crystal structure, modulated by static and dynamic disorder, with a subtle electronic structure requiring methods beyond density functional theory to correctly treat the many-body and relativistic effects. 
As such, the halide perovskites represent a challenge to predictive materials modelling, in a system of great experimental interest, and where there is considerable motivation to improve on the status quo. 

\begin{figure*}
\includegraphics[]{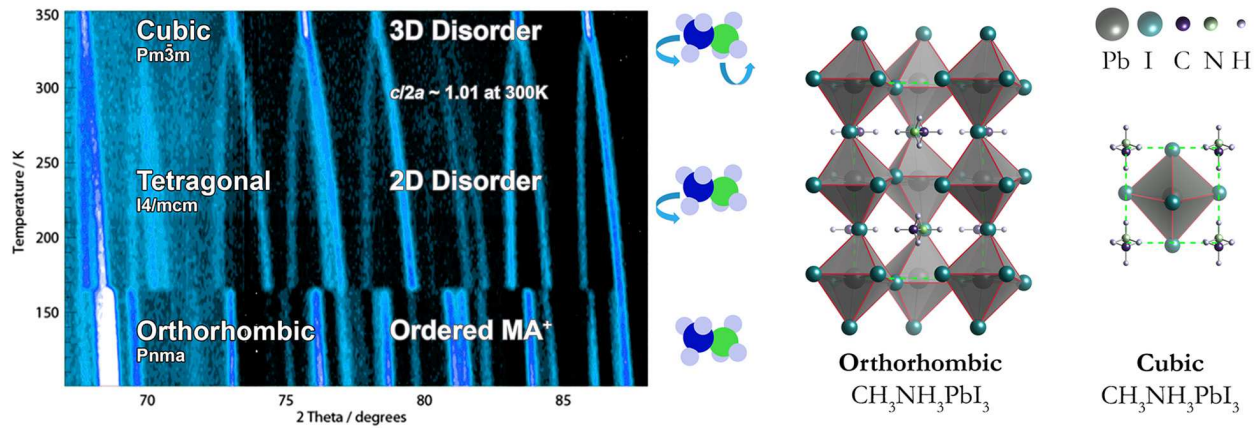}
\caption{
The high-resolution powder neutron diffraction pattern of the hybrid halide perovskite \ce{CH3NH3PbI3} is shown in the left panel (adapted with permission from Ref. \onlinecite{Frost2016a} based on data in Ref. \onlinecite{Weller2015}). This illustrates the low and high temperature phase transitions. While an ordered \ce{CH3NH3+} sub-lattice is expected in the orthorhombic phase, orientational disorder increases with temperature. 
The crystallographic unit cells of the pseudo-cubic and orthorhombic perovskite phases are shown in the right panel (adapted with permission from Ref. \onlinecite{Brivio2015a}). The associated structure files can be accessed from \url{https://github.com/WMD-group/hybrid-perovskites}.
}
\label{fig1}
\end{figure*}

\section{Crystal Structures and Lattice Dynamics} 

\subsection{Phase diversity}
(Hybrid) perovskites of the type \ce{ABX3} form a crystal structure with an (organic) A site cation contained within an inorganic framework \ce{BX3} of corner sharing octahedra. 
Halide substitution on the X site (X = \ce{Cl-}, \ce{Br-}, \ce{I-}), metal substitutions on the B site (B = \ce{Pb^2+}, \ce{Sn^2+}), and cation substitution on the A site (A = \ce{CH3NH3+}, \ce{HC(NH2)2+}, \ce{Cs+}, \ce{Rb+}) lead to varied chemical and physical properties.\cite{Mitzi2001,Mitzi2004a}
In addition to isoelectronic substitutions (e.g. replacing \ce{Pb^2+} by \ce{Sn^2+}), it is possible to perform pairwise substitutions to form double perovskites (e.g. replacing two \ce{Pb^2+} by \ce{Bi^3+} and \ce{Ag^+}).\cite{Savory2016,Volonakis2016}

In the first report of \ce{CH3NH3PbI3} by Weber in 1978, the crystal structure was assigned as cubic perovskite (space group $Pm\bar{3}m$).\cite{Weber1978,Weber1978a}
The anionic \ce{PbI3-} network is charge balanced by the  \ce{CH3NH3+} molecular cation.
The symmetry of \ce{CH3NH3+}  ($C_{3v}$) is incompatible with the space group symmetry ($O_h$) unless orientation disorder (static or dynamic) is present.
The crystal structure solved from X-ray or neutron diffraction data usually spread the molecules over a number of orientations with partial occupancy of the associated lattice sites.
A common feature of perovskites is the existence of phase changes during heating (typically from lower to higher symmetry) as shown in Figure \ref{fig1}. 
In hybrid halide perovskites containing methylammonium, these are orthorhombic ($Pnma$), tetragonal ($I4/mcm$) and cubic ($Pm\bar{3}m$) phases.\cite{Weller2015} 
For \ce{CH3NH3PbI3} the $Pnma$ to $I4/mcm$ phase transition is first-order with an associated discontinuity in physical properties, while the $I4/mcm$ to $Pm\bar{3}m$ phase transition is second-order with a continuous evolution of the structure and properties.\cite{Onoda-Yamamuro1990,Weller2015}

The phase transitions are linked to a change in the tilting pattern of the inorganic octahedral cages, and order-disorder transitions of the molecular sub-lattice.\cite{Onoda-Yamamuro1990,Yamamuro1992a,Onoda-yamamuro1992}
X-ray diffraction (XRD) measurements upon cooling (heating) suggest the inclusion of tetragonal in orthorhombic phases (and vice-versa).\cite{Hutter2016a} 
This is often observed for first-order solid-state phase transitions. 
In addition, it has been suggested that the presence of multiple photoluminsence peaks at low T is due to the coexistence of ordered and disordered orthorhombic domains.\cite{Dar2016}

Similar phase behaviour tends to be seen for other compositions, however the transition temperatures vary.
In \ce{CH3NH3PbI3} the orthorhombic to tetragonal transition temperature is $162$ K, becoming cubic by around $328$ K.
\ce{CH3NH3PbBr3} is cubic above $237$ K.\cite{Poglitsch1987c} 
In addition, compounds such as \ce{HC(NH2)2PbI3} (FAPI) and \ce{CsSnI3} feature phase competition between a corner-sharing octahedra perovskite phase (black in appearance) and edge-sharing octahedra molecular crystals (yellow or white in appearance).\cite{Weller2015b}
Only the corner-sharing perovskite phase is of interest for solar energy applications. 

\begin{figure*}
\includegraphics[]{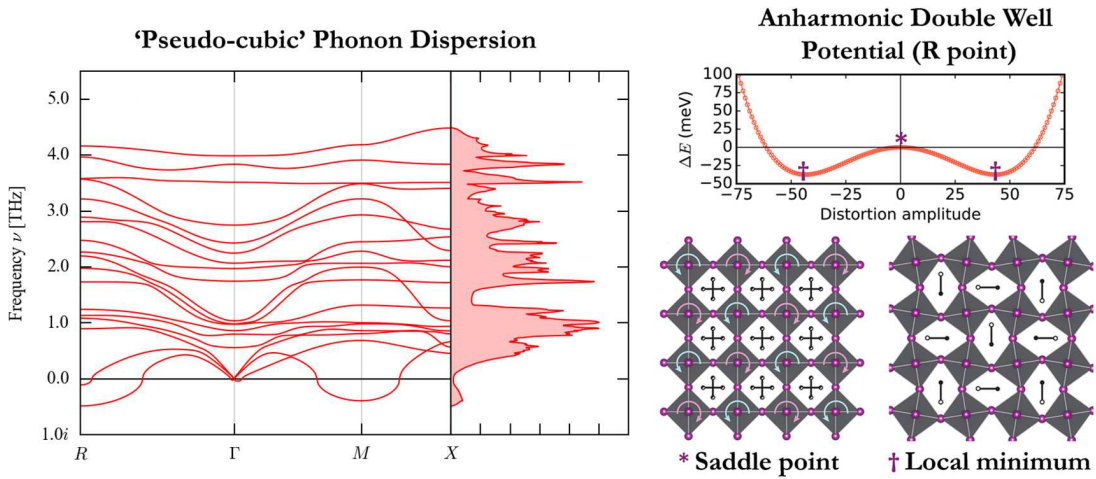}
\caption{
    (Left) The harmonic phonon dispersion for \ce{CH3NH3PbI3} from a `pseudo-cubic' structure. 
    The imaginary frequencies of acoustic modes at the $M$ ($q=\frac{1}{2},\frac{1}{2},0$) and $R$ ($q = \frac{1}{2}, \frac{1}{2}, \frac{1}{2}$) Brillouin zone boundary points correspond to an instability expressible in a supercell as alternate tilting of the octahedra.
    (Right) Following the imaginary acoustic mode at the $R$ point in a $2\times2\times2$ supercell expansion shows a double-well potential in the DFT internal energy. 
    The saddle point corresponds to a $1\times1\times1$ cubic structure, whilst the two local minima correspond to a distorted structure of lower symmetry. 
The energy barrier is small enough to allow both minima can be accessed at room temperature, so the system is expected to exhibit dynamic rather than static disorder. 
Similar behaviour is found at the $M$ point. 
The figure is adapted with permission from Refs. \onlinecite{Whalley2016} and \onlinecite{Beecher2016a}.
The underlying phonon data is available from \url{https://github.com/WMD-group/Phonons}.
}
\label{fig2}
\end{figure*}

\subsection{Local and average crystal environment} \label{localaverage}

The first electronic structure calculation of hybrid halide perovskites was by Chang, Park and Matsuishi in 2004, \cite{Chang2004}
in the local density approximation (LDA) of density functional theory (DFT).
They modelled a static structure where the \ce{CH3NH3+} molecule was aligned along $\langle100\rangle$ (towards the face of the corner-sharing \ce{PbI3-} framework), but found that the barrier for rotation to $\langle111\rangle$ was less than $10$ meV. 
This small barrier for cation rotation gave credence to a prior model that the molecular sub-lattice was dynamically disordered.\cite{Poglitsch1987c}
Similar barriers were later found within the generalised gradient approximation (GGA) of DFT.\cite{Brivio2013} 

\emph{Ab initio} molecular dynamics (MD), neutron scattering\cite{Leguy2015b,Chen2015s} and time-resolved infra-red\cite{Bakulin2015a} data all indicate a 1--10 picosecond reorientation process of the molecular cation at room temperature.
As a result of (by definition) anharmonic molecular rotation, and large-scale dynamic distortions along soft 
vibrational modes, the local structure can deviate considerably from that sampled by diffraction techniques. Bragg scattering does not probe local disorder, if it preserves long-range order on average.
Knowledge of these locally broken symmetries is essential for meaningful electronic structure calculations, where the broken symmetry results in a lifting of degeneracy, and a potentially quite different solution. 

In spite of the larger cation, FAPI appears to possess a similar timescale of rotation to MAPI\cite{Weller2015b}. 
A lighter halide (and therefore smaller cage) results in faster rotation, in spite of the greater steric hindrance.\cite{selig2017organic}
Together, these data suggest that the molecular rotation is a function of the local inorganic cage tilting. The relatively insignificant mass of the organic cation follows the distortion of the cavity. 

Spontaneous distortions can also be observed in the vibrational spectra.
The calculated harmonic phonon dispersion for MAPI in the cubic phase is presented in Figure \ref{fig2}.
The acoustic phonon modes soften as they approach the $M$ ($q = \frac{1}{2}, \frac{1}{2}, 0$) and $R$ ($q = \frac{1}{2}, \frac{1}{2}, \frac{1}{2}$) Brillouin zone boundary points. 
This zone boundary instability can only be realised in an even supercell expansion, where it corresponds to anti-phase tilting between successive unit cells.
This behaviour is characteristic of the perovskite structure, and can be described by the Glazer tilt notation.\cite{Glazer1972,Woodward1997} 

Within the frozen-phonon approximation the potential energy surface can be traced along the soft acoustic $M$ and $R$ phonon modes. 
In both cases this results in a double well with an energy barrier $\sim k_BT$ at the saddle point.\cite{Whalley2016}  
At room temperature the structure is dynamically disordered, with continuous tilting.
The structure is locally non-cubic but possess only cubic Bragg scattering peaks.\cite{Beecher2016}
Indeed, MD simulations of halide perovskites show continuous tilting of octahedra at room temperature.\cite{Frost2014,Quarti2015a,Weller2015b}
As temperature decreases, the structural instability condenses via the soft mode at the $R$ point (with an energy barrier of 37 meV) into the lower symmetry tetragonal phase. 
This is followed by condensation of the $M$ point (with an energy barrier of 19 meV) to the orthorhombic phase.\cite{Whalley2016}
Whilst the molecular cation continuously rotates with the inorganic tilts in the cubic phase, and is partly hindered in the tetragonal phase, it can only librate in the low temperature orthorhombic phase.

In the static picture (as in the case of an electronic band structure calculated for a single ionic snapshot),
the organic cation plays no direct role in optoelectronic properties of the material as the molecular electronic levels lie below that of the inorganic framework.
Allowing motion, the electrostatic and steric interaction between the organic molecule and inorganic framework couples tilting and distortion of the octahedra to the organic cation motion.
These tilts and distortions vary the atomic orbital overlap, perturbing the band-structure and bandgap.\cite{Mosconi2014,Quarti2015a,Whalley2016,Saidi2016}
The electronic structure thus becomes sensitive to temperature, which will be discussed further in Section II.

\subsection{Thermodynamic and kinetic stability}

\textit{Ab initio} thermodynamics has emerged as a powerful tool in materials modelling, with the ability to assess the stability of new materials and place them on equilibrium phase diagrams even before experimental data is available. \cite{Reuter2003,Kim2012,Jackson2015a}
The total energy from DFT calculations approximates the internal energy of the system. 
By including lattice vibration (phonon) and thermal expansion contributions, the Gibbs free energy 
and other thermodynamic derivatives can be evaluated.\cite{Stoffel2010}
In the context of photovoltaic materials, this has been applied to \ce{Cu2ZnSnS4} 
and used to identify the processing window where a single-phase compound can be grown in equilibrium.\cite{Jackson2014}
For the tin sulfide system it shows the close competition between SnS, \ce{SnS2} and \ce{Sn2S3}.\cite{skelton2017chemical}

An issue with hybrid perovskites and other metal-organic frameworks is that the calculated heat of formation is close to zero.
The decomposition reaction 
\begin{equation}
\ce{CH3NH3PbI3} \rightarrow \ce{CH3NH3I} + \ce{PbI2}
\end{equation}
has been predicted to be exothermic.\cite{Zhang2015k}
Subsequent calorimetric experiments have supported the prediction that hybrid lead halide perovskites are metastable.\cite{Nagabhushana2016}
It is likely that these materials are only formed due to entropic (configurational, vibrational and rotational) contributions
to the free energy.

The concept of metastable materials is attracting significant interest.\cite{Caskey2014,Walsh2015b,sun2016thermodynamic,Skelton2017} 
These are materials that do not appear on an equilibrium phase diagram but can be synthesised with a finite (useful) lifetime.
For such materials, the chemical kinetics become critical,
and formation and stability and can be particularly sensitive to local gradients in chemical potential (e.g. compositional, thermal, electronic). 
Though kinetic factors can be calculated with first-principles techniques, this is a more cumbersome and costly process than equilibrium bulk thermodynamics, which requires only total energies of local minimum structures. 
To our knowledge, there have been no rigorous attempts to model the kinetics of decomposition pathways for hybrid perovskites over complete chemical reactions. 

\begin{figure*} \label{f3}
\includegraphics[]{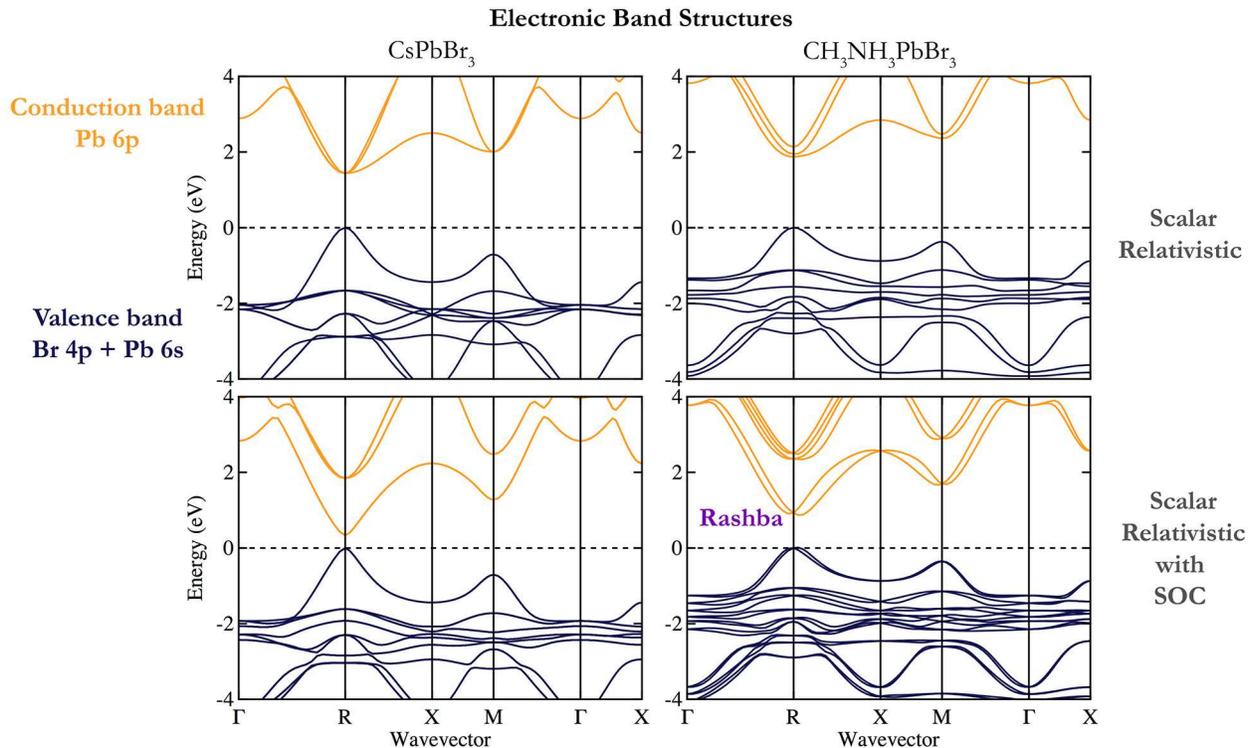}
\caption{
The electronic band structures of the inorganic perovskite \ce{CsPbBr3} and hybrid perovskite \ce{CH3NH3PbBr3} in the cubic phase.
    One effect of the organic cation is to widen the bandgap located at the $R$ point due to the larger lattice constant. 
    Spin-orbit coupling reduces the bandgap in both materials. 
The presence of \ce{CH3NH3+} in the hybrid perovskite results in a non-centrosymmetric crystal, with an associated relativistic Rashba-Dresselhaus splitting of the lower conduction band.
While labels of the special points are those of the cubic perovskite structure (space group $Pm\bar{3}m$),  
    the static model of the hybrid perovskite formally has  \textit{P1} symmetry . 
Points equivalent for a cubic crystal (e.g. $M=\frac{1}{2},\frac{1}{2},0$; $M'=0, \frac{1}{2},\frac{1}{2}$;  $M''=\frac{1}{2},0,\frac{1}{2}$) are inequivalent here.
}
\label{fig3}
\end{figure*}

\subsection{Anharmonic lattice vibrations and thermal conductivity}

Electronic structure theory is most often carried out in the Born-Oppenheimer approximation where the nuclei are static classical point charges. 
To consider thermal vibrations, expansion, or heat flow, the theoretical framework of lattice dynamics can be used.\cite{Stoffel2010} 
 
In the harmonic approximation, the small-perturbation lattice dynamics are fully specified by second-order force-constants of individual atoms. 
These are readily constructed into the so-called dynamical matrix. 
The eigenstates of this matrix are the normal modes of vibration with an associated frequency. 
The description of collective vibrational excitations in crystals can be simplified with second quantization to the creation and annihilation of phonon quasiparticles, specified by these normal modes.
Thermal expansion coefficients, system anharmonicity (e.g. modal \grun parameters) and the temperature-dependence of other properties can be calculated in the quasi-harmonic approximation (QHA). 
In this formalism, the lattice dynamics is harmonic at a given temperature; however, the cell volume is scaled by thermal expansion to account for finite-temperature anharmonic effects. 

The thermal expansion coefficient of MAPI in the cubic phase has been calculated with the QHA.
The value is sensitive to the exchange-correlation functional used.
For example, a value of $3.0 \times 10^{-5} /K$ is calculated with the PBE functional with Tkatchenko-Scheffler dispersion corrections.\cite{Saidi2016b} 
The PBEsol functional produces a value of $12.5 \times 10^{-5} /K$.\cite{Brivio2015a} 
These compare to finite temperature scattering measures of $1.9 \times 10^{-5} /K$ by X-ray,\cite{Baikie2013} and $13.2 \times 10^{-5} /K$ by neutron diffraction.\cite{Weller2015} 
Even taking the smallest value above, the expansion coefficient is one order of magnitude greater than silicon.\cite{Madelung2003} 
This highlights the strong deviation from harmonic behaviour in halide perovskites.

In the harmonic approximation (and similarly the QHA), the eigenmodes 
of the dynamical matrix are orthogonal and the resulting phonons are non-interacting.
Consequently phonon lifetimes are infinite as the phonons do not scatter; thermal conductivity is ill-defined. 
To calculate phonon-phonon scattering, and so its contribution to finite thermal conductivity, anharmonic lattice dynamics need to be considered.
A computational route is to use perturbative many-body expansion, e.g. as implemented in \textsc{PHONO3PY},\cite{Togo2015} which includes third-order force constants. 
For \ce{CH3NH3PbI3}, 41,544 force evaluations are required to evaluate the third-order force constants, compared to 72 for second-order (harmonic) force constants.\cite{Whalley2016}
Consequently, these calculations are vastly more computationally expensive. 
Using this approach, phonon-phonon scattering rates are calculated to be three times larger in MAPI compared to standard covalent semiconductors \ce{CdTe} and \ce{GaAs}.\cite{Whalley2016} 
The phonons barely exist for a full oscillation before they split or combine into another state. 
Consequently, mean free paths are on the nanometer rather than more typical micrometer scale. 
Lattice thermal conductivity is extremely low, 0.05 Wm$^{-1}$K$^{-1}$ at 300 K.\cite{Whalley2016}
This combination of high electrical and low thermal conductivity makes these compounds potential thermoelectrics.\cite{He2014,Mettan2015}

In highly anharmonic systems perturbatively including third-order force constants may not be sufficient
to describe the true dynamics. 
Yet going further in the lattice dynamics formalism becomes prohibitive.
Besides, it is not obvious whether the fundamental tenant of lattice dynamics, of expanding in small displacements around a minimum structure, is correct for such soft and highly anharmonic materials. 
In contrast, MD treats anharmonic contributions to all orders, but as it stochastically explores the phase space, long integration times are required to sample rare events. 
Finite size effects also mean that only phonon modes commensurate with the supercell are sampled,
so size convergence has to be carefully considered. 

Classical interatomic potentials derived from first-principles calculations have been developed for hybrid perovskites.\cite{Mattoni2015a,Hata2016,handley2017new} 
Such models are able to correctly reproduce crystal structures, as well as mechanical and vibrational properties. 
Calculation of thermal conductivity from molecular dynamics simulated for MAPI predicted values of 0.3 to 0.8 Wm$^{-1}$K$^{-1}$ at 300 K.\cite{Wang2016d,Caddeo2016} 
Although still ultra-low, these values are greater than that values calculated by perturbation theory. 
It should be noted that these values give upper limits for thermal conductivity as they refer to a defect-free isotopically-pure bulk sample.

\section{Electronic Structure}

Despite the dynamic disorder just discussed, in many respects halide perovskites display characteristics of traditional inorganic semiconductors, with a well-defined electronic band structure and electron/hole dispersion relations.
However, subtleties emerge upon closer examination, when the electronic structure is correctly modelled. 

\subsection{Many-body and relativistic effects} \label{Mbre}

Perhaps surprisingly, local and semi-local exchange-correlation functionals 
provide a reasonable estimate for the bandgaps of these heavy metal halide materials.
This is due to a cancellation of errors. 
For Pb-based perovskites, the conduction band has mainly Pb 6p character. 
Due to the large nuclear charge, the electronic kinetic energy requires a relativistic
treatment, and spin-orbit coupling (SOC) becomes significant. 
The first-order effect is a reduction in bandgap by as much as 1 eV\cite{Brivio2014a}, as the degenerate 6p orbitals are split and moved apart in energy. 
This is shown in Figure \ref{fig3} for the bromide compounds.
The typical bandgap underestimation of GGA functionals is offset by the absence of relativistic renormalisation.

SOC is not expected to have a large impact on the structural properties of the Pb-based compounds as the (empty) conduction band is mainly affected.
By the Hellmann-Feyman theorem the force on atoms depends only on the electron density, which is provided by the occupied orbitals. 
Accurate force-constants (as needed in both molecular and lattice dynamics) can be calculated without SOC considerations.\cite{PerezOsorio2015a}

There have been a number of electronic structure calculations considering many-body interactions beyond DFT.\cite{Brivio2014a,Filip2014c,Umari2014}
Quasi-particle self-consistent \textit{GW} theory shows that the band dispersion (and so density of states, optical character and effective mass) is considerably affected by both the \textit{GW} electron correlation and spin-orbit coupling.\cite{Brivio2014a}
Some materials see only a rigid shift of band structure (retaining DFT dispersion relations),\cite{VanSchilfgaarde2006,Butler2016} but this is not the case for hybrid perovskites.
This point has not been fully appreciated, in part because DFT codes are more widespread and convenient to generate data.

A consequence of SOC when combined with a local electric field is the Rashba-Dresselhaus effect, a splitting of bands in momentum space.\cite{Kepenekian2015} 
This can be understood as an electromagnetic effect, where the magnetic moment (spin) of the electron interacts with a local electric field, to give rise to a force which displaces it in momentum space. 
Up and down spins are displaced in opposite directions, and this displacement is a function (in both size and direction) of the local electric field, which will depend on the local dynamic order. 
For a static structure, this is demonstrated in Figure \ref{fig3} for \ce{CH3NH3PbBr3}. 
Neglecting SOC, the cubic phase has band extrema at the $R$ point (a direct bandgap).
With SOC the valence and conduction band each split into valleys symmetrical around $R$.
The splitting is much more pronounced in the \ce{Pb} $6p$ conduction band (compared to the \ce{Br} $4p$ valence band), as expected from the $Z^4$ dependence of spin-orbit coupling.
This asymmetry in the band extrema results in direct-gap like absorption and indirect-gap like radiative recombination, which we discuss later.

The relativistic spin-splitting can only occur in crystals that lack a centre of inversion symmetry, a prerequisite for generating a local electric field. 
The cubic representation of \ce{CsPbBr3} has an inversion centre, so while SOC affects the bandgap through the separation of Pb 6p into p$_{\frac{1}{2}}$ and  p$_{\frac{3}{2}}$ combinations,
no splitting of the band extrema away from the high symmetry points is observed (see Figure \ref{fig3}).
This is true only for a static cubic structure. 
As discussed earlier, hybrid halides will have continuous local symmetry breaking. 
Calculations based on static high symmetry structures are not representative
of the real (dynamic) system and can be misleading. 


The calculation of electronic and optical levels associated with intrinsic and extrinsic point defects 
 will be particularly sensitive to the electronic structure method used.
Neglect of SOC and self-interaction errors can result in an incorrect position of the valence or conduction band edges, thus introducing spurious errors in defect energy levels and predicted defect concentrations.
Du\cite{Du2015} showed how for the case of an iodine vacancy, a deep (0/+) donor level is predicted for GGA-noSOC,
while a resonant donor level is predicted for GGA-SOC and HSE-SOC treatments of electron-exchange and correlation.

\subsection{Electron-phonon coupling}

Going beyond the Born-Oppenheimer approximation with perturbation theory, we can consider the interaction of the electronic structure with vibrations of the lattice. 
Electron-phonon coupling can perturb the electronic band energies (changing the bandgap), and couple electronic excitations (the hole and electron quasi-particles) into vibrational excitations (phonon quasi-particles). 
In a semiconductor, charge carrier scattering is often dominated by this electron-phonon interaction. 
The strength of these processes can set a limiting value on mobility. 
Electron-phonon coupling is often calculated in a second-order density functional perturbation theory calculated for a static (rigid ion) structure.
For normal covalent systems, this term is expected to dominate over the first order contribution from the acoustic deformation potential as vibrations are typically small. 
These calculations are difficult to converge, as integration is over both electronic and vibrational reciprocal space, and the electron-phonon interaction is often found to be a non-smooth function.\cite{RevModPhys.89.015003}

In recent work\cite{Whalley2016} we developed a method to calculate the electron-phonon interaction of soft anharmonic phonon modes, and applied this to the acoustic zone boundary tilting in hybrid halide perovskites described earlier. 
We solve a one-dimensional Schr\"odinger equation for the nuclear degree of freedom along the phonon mode, and then combine the resulting thermalised probability density function (which includes zero point fluctuations and quantum tunnelling) with a bandgap deformation potential along this mode. 
The method includes quantum nuclear motion, goes beyond the harmonic regime, but only contains the first-order contribution to the electron-phonon coupling of the bandgap deformation. 
A positive bandgap shift of 36 meV (\textit{R} point phonon) and 28 meV (\textit{M} point phonon) 
is predicted at T = 300 K.
Saidi et al. sampled all non-soft harmonic phonons using a Monte Carlo technique,\cite{Saidi2016} finding significant differences with (more standard) perturbation theory results.
Electron-phonon interactions can be calculated with MD, but as with phonon-phonon scattering, 
achieving convergence with respect to electronic (\textit{k}-point sampling and basis set) 
and vibrational ($q$-point sampling and supercell size), while maintaining sufficient integration time to capture rare processes, is costly.

Recently a `one shot' method has been developed to calculate bandgap renormalization and phonon-assisted optical absorption, and applied to \ce{Si} and \ce{GaAs}.\cite{Zacharias2016} 
Nuclei positions are carefully chosen as a representative sample from the thermodynamic ensemble, and the electronic structure is needed for this static structure only---a significant increase in computational efficiency. 
Such techniques may provide a promising method to calculate the electron-phonon coupling of complex materials, but so-far are only valid in the harmonic phonon approximation. 
They have not yet been tested for the family of hybrid halide perovskites or other more complicated crystal structures.

\subsection{Charge carrier transport}

We now consider some aspects of charge carrier transport in hybrid halide perovskites.
The minority-carrier diffusion length is the average length a photo-excited (or electronically-injected) carrier travels before recombining. 
In a working photovoltaic device, the diffusion length must be sufficient for photo-generated charges to reach the contacts.
The minority-carrier diffusion length is a product of the diffusivity $D$ and lifetime $\tau$ of minority charge carriers, $L_d = \sqrt{D\tau}$.

Minority-carrier diffusion lengths in MAPI are reported to be considerably longer than other solution processed semiconductors.\cite{Li2015zz}

Long lifetimes (large $\tau$) can be partly attributed to the `defect-tolerance' of hybrid perovsites (discussed in Section \ref{defects}), reducing the rate of ionised-impurity scattering and non-radiative recombination.  

The effective mass of both electrons and holes in hybrid halide perovskites is small (though careful calculations including spin-orbit coupling indicate that the band extrema do not show a paraoblic dispersion relation, and so the concept of effective mass is ill-defined\cite{Brivio2014a}).
Given effective masses of $< 0.2 m_e$,  the carrier mobility of MAPI ($< 100$ \mob) is modest in comparison to conventional semiconductors such as \ce{Si} or \ce{GaAs} ($> 1000$ \mob).\cite{Stranks2015b}
Carrier mobility must be limited by strong scattering.

Low temperature mobility in this material reduces as a function of temperature as T$^{-1.5}$, which provides circumstantial evidence for being limited by acoustic phonon scattering.\cite{Karakus2015,Yi2016a}
However, if we only consider acoustic phonon scattering (which is elastic due to the population of acoustic modes), the calculated mobility is orders of magnitude larger than experiment. 
A key realisation is that the soft nature of these semiconductors results in optical phonon modes (see Figure \ref{fig2}) below thermal energy.\cite{Brivio2015a,PerezOsorio2015a}
Optical phonon scattering is inelastic and dominates once the charge carriers have sufficient energy to generate the phonon modes.\cite{Leguy2016} 
Through solving the Boltzmann transport equation parameterised by DFT calculations, at room temperature the scattering from longitudinal optical phonons is most relevant in limiting mobility.\cite{Wright2016,Filippetti2016}

Carrier mobility will be further limited by scattering from point and extended lattice defects.\cite{Ball2016}
Fluctuations in electrostatic potential resulting from dynamic disorder provide a macroscopic structure from which carriers will also scatter.\cite{Frost2014,Ma2014d}

\begin{figure*}
\includegraphics[]{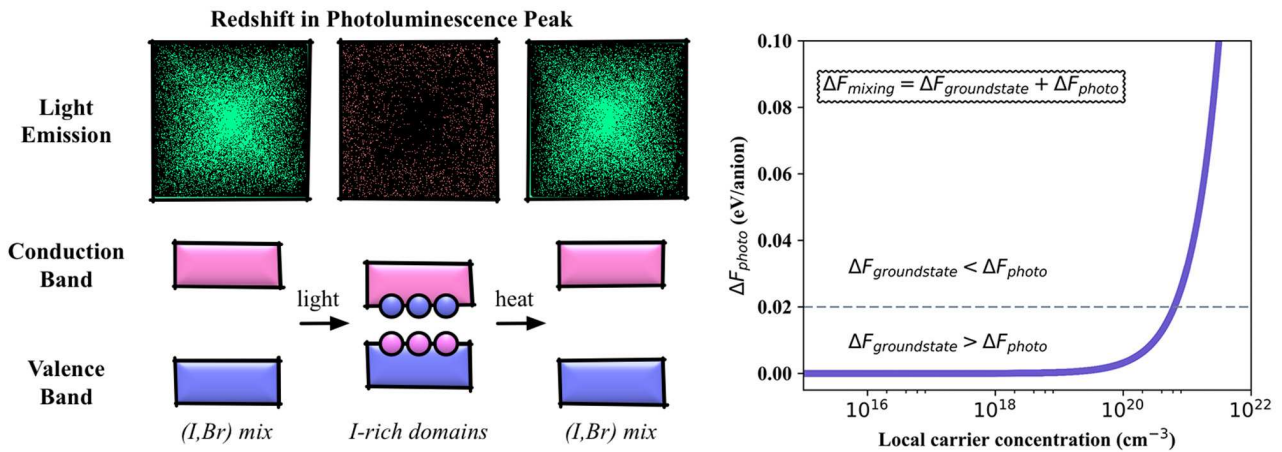}
\caption{
Ion transport occurs in halide perovskites: they are mixed ionic-electronic conductors. The vacancy-mediated diffusion of halide anions has been associated with both current-voltage hysteresis of solar cells and the rapid interchange between iodide, bromide and chloride materials.
One point of controversy remains the reversible ion segregation observed in mixed (Br,I) systems. 
Alloyed materials have been found to phase separate upon illumination, but recover their initial state when the light source is removed. 
The phase separation is associated with a striking red-shift in  photoluminesence spectra.
A statistical mechanical analysis of ground-state DFT calculations suggested a large miscibility gap \cite{Brivio2016}, while the charge carriers generated upon illumination can provide an additional driving force for phase separation.\cite{Slotcavage2016}
The results from a simple thermodynamic model are shown in the right panel, where the free energy of mixing contains contributions from the groundstate  ($\Delta F_{groundstate}$) with an additional component due to the difference in bandgaps between the mixed (I,Br) and phase separated I-rich phases ($\Delta F_{photo}$). 
The latter contribution requires local carrier concentrations approaching 10$^{21}$ cm$^{-3}$ to make a substantial contribution to the overall mixing energy.
}
\label{fig4}
\end{figure*}

\section{Photophysics and Solar Cells}

Recent research interest in hybrid halide perovskites is mainly due to their use as the active layer in efficient solar cells.
There are areas of the underlying physics which are not yet developed, and which may be limiting progress in the field.
Ion migration is poorly understood and has been correlated with hysteresis effects\cite{Eames2015a,Richardson2016} and device degradation.
Defects which act as recombination centres have not been identified and characterised. 
Additionally, interfaces have not been optimised for optimal charge carrier extraction.
We outline these issues -- where theory and simulation have much to contribute -- in the following section.

\subsection{Ion migration} 

Charged point defects in the bulk allow for mass transport of ions and can result in spatial fluctuations of electrostatic potential.
For solid-state diffusion to be appreciable in magnitude, there needs to be a high concentration of defects and a low activation energy for diffusion. 

The equilibrium concentration of charged vacancy defects is calculated as being in excess of 0.4\% at room temperature in MAPI.\cite{Walsh2015}
Low defect formation energies and free-carrier concentrations found across the hybrid halide perovskites indicate that Schottky defects are prevalent across this family of materials.
While each point defect is charged, they are formed in neutral combinations so that a high concentration of lattice vacancies does not require a high concentrations of electrons or holes to provide charge compensation. 

The ion migration rate is given by:
\begin{equation}
\Gamma = \nu \exp \left( \frac{-\Delta H^{\mathrm{diff}}}{k_BT} \right)
\end{equation}
where $\Delta H^{\mathrm{diff}}$ is the activation energy for solid-state diffusion,
and $\nu$ is the attempt frequency. 
In MAPI the diffusion of methylammonium cations, iodide anions and protons have been considered in the literature.\cite{Eames2015a,Egger2015c,Azpiroz2015a}
Activation energies calculated from first principles show that the predominant mechanism for ion migration is the vacancy assisted hopping of iodide ions.\cite{Eames2015a}

Based on a bulk activation energy of 0.58 eV\cite{Eames2015a}, a rate of 733 hops per second would be expected at T = 300 K, with an associated diffusion coefficient of 10$^{-12}$cm$^{2}$s$^{-1}$.
Effective activation energies as low as 0.1 eV have been reported experimentally,\cite{Bryant2015,Game2017} which likely 
correspond to diffusion along extended defects (dislocations, grain boundaries, surfaces)\cite{Shao2016a,Yun2016}.
The corresponding diffusion rate of 10$^{-5}$cm$^{2}$s$^{-1}$ is very fast, but comparable to surface 
diffusion of iodine observed in other compounds.\cite{Chandra1980}
It is also comparable to the diffusion coefficient of 4$\times 10^{-6}$cm$^{2}$s$^{-1}$ predicted by classical molecular dynamics. \cite{Delugas2016}

Modelling ion diffusion at device scales is not yet possible with \emph{ab-initio} methods.
Parametrised drift-diffusion modelling of ion and electron density indicate that slow moving ions can explain the slow device hysteresis.\cite{VanReenen2015,Richardson2016} 
A vacancy diffusion coefficient of the order of 10$^{-12}$cm$^2$s$^{-1}$ is consistent with both predictions and transient measurements.\cite{Eames2015a}

It has been suggested that ion migration within mixed-halide compositions is the result of a non-equilibrium process induced by photo-excitation.
X-ray diffraction measurements by Hoke et al.\cite{Hoke2015} show that under illumination the mixed halide perovskite $\ce{MAPb(I_{1-x}Br_x)3}$ segregates into two crystalline phases: one iodide-rich and the other bromide-rich.
This segregation leads to reduced photovoltaic performance via charge carrier trapping at the iodide-rich regions.
In some reports, after a few minutes in the dark the initial single phase XRD patterns are recovered. 
This reversible process is unusual and defies the common assumption made that ion and electron transport are decoupled.

A schematic outlining the phase segregation process is shown in Figure \ref{fig4}.
A phase diagram constructed from first-principles thermodynamics found a miscibility gap for a range of stoichometries at room temperature.\cite{Brivio2016}
This suggests that a mixed-halide material is metastable and will phase segregate after being excited by light, which follow a decreasing free energy gradient towards halide-rich areas formed prior to light excitation (such as grain boundaries).
The accumulation of charge carriers increases lattice strain and drives further halide segregation.  
Our calculations indicate that the transition between mixing and segregation will occur at a local carrier concentration of $10^{21}$ cm$^{-3}$, which would require accumulation into small regions of the material. 

\subsection{Electron-hole recombination} \label{EHR}

The open-circuit voltage (V$_{\mathrm{OC}}$) of a solar cell is determined by the rate of charge carrier recombination in the material, as no photogenerated charges are being extracted and so all are recombining. 
When operated to generate power, the rate of recombination competes with the rate of charge extraction, limiting the fill factor of the solar cell. 
Combined, rates of recombination specify the photovoltaic potential of a material. 

Recombination is usually separated into three channels: 
non-radiative; radiative; and Auger. 
These respectively correspond to: one; two; and three electron processes.
Assuming that the prefactors for the rates of these processes are constant, the carrier density in an intrinsic semiconductor can be modelled as a rate equation:
\begin{equation}
\frac{ \mathrm{d}n}{\mathrm{d} t} = G - nA - n^2B - n^3C
\end{equation}
where $G$ is the rate of electron-hole generation, $n$ the density of charge-carriers.

While non-radiative recombination is limiting in many inorganic thin-film technologies, hybrid perovskites are not significantly affected. 
This is surprising for the high density of defects expected
for a material deposited from solution at relatively low temperatures, leading to the material being described as `defect tolerant'.\cite{Berry2016} 

%

Radiative (bimolecular) recombination is slower than would be expected for a direct bandgap semiconductor. 
Recent calculations\cite{Azarhoosh2016, Zheng2015} revealed how relativistic Rashba splitting can suppress radiative recombination at an illumination intensity relevant to an operating solar cell.
After photoexcitation, electrons thermalise to Rashba pockets in the conduction band minima away from the high symmetry point in reciprocal space.
This leads to an indirect charge recombination pathway as the overlap in $k$-space between occupied states near upper valence and lower conduction bands diminishes.
It has also been suggested that direct recombination is suppressed due to the pockets of minima being spin-protected.\cite{Zheng2015}
Direct gap radiative recombination is reduced by a factor of 350 at solar fluences, as electrons must thermally repopulate back to the direct gap.\cite{Azarhoosh2016}
This is in agreement with the temperature-dependence of the bimolecular rate measured experimentally \cite{Hutter2016a} and calls into question the validity of models where a global radiative recombination rate independent of carrier concentration is used.

Auger recombination is only significant at fluences well above solar radiation, but is important for understanding laser photophysics. 

Ferroelectric effects could contribute to electron-hole separation due to electrostatic potential fluctuations in real space.
Although the molecular cation plays no direct role in charge generation or separation it could have a part to play in charge transport through the formation of polar domains.\cite{Frost2014b,Ma2014d}
Macroscopic ferroelectric order is not necessary to explain device behaviour in a 3D drift-diffusion simulation.\cite{Sherkar2015}
A multiscale Monte Carlo code based on a model Hamiltonian parameterized for the inter-molecular dipole interaction in MAPI, explored the results of this dynamic polarisation.\cite{Frost2014}
This predicts the formation of antiferroelectric domains which minimise energy via dipole-dipole interaction, which work against a cage-strain term preferring ferroelectric alignment.\cite{Leguy2015b}
This provides electrostatically preferred pathways for electrons and holes to conduct.
Developing more accurate models and measurements of the nature and effects of lattice polarisation in these materials is the subject of on-going research efforts.


\subsection{Defect levels in the bandgap}\label{defects}

To understand why the rate of non-radiative recombination is low we consider the known defect properties of hybrid perovskites.
Defects appear to have a minimal impact upon charge carrier mobility and lifetime,\cite{Brandt2015a}
which can be attributed to a combination of large dielectric constants and weak heteropolar bonding. 

Under the Shockley-Read-Hall model for semiconductor statistics non-radiative recombination is mediated through deep defect states in the gap.\cite{PhysRev.87.835}
Shallow defect states can act as traps but the carriers are thermally released to the band before recombination can occur.
Hybrid perovskites -- with high dielectric constant and low effective mass -- show a tendency towards benign shallow defects under the hydrogenic model:\cite{Yu1996}
\begin{equation} \label{hydeqn}
E_n = - \frac{m^*}{m_0}\frac{1}{2n^2\epsilon_0^2}
\end{equation}
where $\frac{m^*}{m_0}$ is the effective mass ratio, $\epsilon_0$ is the static dielectric constant, and $n$ is an integer quantum number for given energy level. Atomic units are used and so energy is given in Hartrees. 

In Table \ref{tab:defectlevels} we give the first hydrogenic defect level for MAPI, Si and CdTe, where the binding energy for MAPI is only 3 meV.
For ionic materials, one would expect a large central cell correction that could result in much deeper levels, for example, as seen for the colour centres in alkali halides.\cite{Stoneham1975}
It was shown numerically that the on-site electrostatic potentials in the I-II-VII$_3$ perovskites are relatively weak owing to the small charge of the ions (e.g \ce{Cs+Pb^{2+}I^-_3}) compared to other perovskite types (e.g. \ce{Sr^{2+}Ti^{4+}O^{2-}_3}),\cite{Brivio2014} which would also support  more shallow levels. 
In addition, arguments based on covalencey have also been proposed.\cite{Brandt2015a}

\begin{table}
\caption{\label{tab:defectlevels}The first shallow donor defect level in \ce{MAPI}, \ce{Si} and \ce{CdTe} calculated from effective mass theory using Equation \ref{hydeqn}. The dielectric constant $\epsilon_0$ can be considered an important descriptor for photovoltaic materials as several important properties (e.g. rate of impurity scattering) scale with its square.}
\begin{ruledtabular}
\begin{tabular}{cccc}
Material & $\frac{m^*}{m_0}$ & $\epsilon_0$ & $E_1 (\mathrm{meV})$ \\
\hline
\ce{CH3NH3PbI3} & 0.15\cite{Frost2014b} & 25.7\cite{Frost2014b} & 3 \\
Si & 0.45\cite{Hava2007} & 11.7\cite{Hava2007} &  45 \\
CdTe & 0.11\cite{Wang2007} & 10.2\cite{Madelung2003} & 14 \\
\end{tabular}
\end{ruledtabular}
\end{table}

\subsection{Beyond the bulk: surfaces, grain boundaries and interfaces}

As perovskite solar cells approach commercial viability,\cite{Park2016} there are considerations to be made beyond the bulk materials.
Surfaces, grain boundaries and interfaces will influence device performance and long-term stability, and become increasingly important as the science is scaled up from lab to production line. 
Accurate interface modelling requires consideration of halide migration, ion accumulation, charge carrier transport and charge carrier recombination at the defect states.
There has been preliminary work, that provides insights, but real systems offer much deeper complexity. 

Perovskite films fabricated through solution processing methods are multicrystalline and so the formation of grain boundaries is inevitable.
The resulting microstructure provides pathways for ion conduction, electron-hole separation and recombination.
The shallow traps introduced are evidenced through improved device performance with increasing grain size\cite{Chen2016} and their thermal activation. 
Calculations suggest that grain boundaries do not introduce deep defects and consequently have negligible effect upon the rate of non-radiative recombination.\cite{Yin2015b, Guo2017}
This is in conflict with spatially resolved photoluminescence\cite{deQuilettes2015a} and cathodoluminescence\cite{Bischak2015a} measurements which evidence greater non-radiative loss at grain boundaries.

Recent calculations using nonadiabatic MD and time-domain density DFT\cite{Long2016a} indicate that grain boundaries localize the electron and hole wavefunctions and provide additional phonon modes.
This leads to increased electron-phonon coupling which in turn will give a higher rate of non-radiative recombination. 

The typical device structure for a perovskite cell is the perovskite absorber layer sandwiched between an electron transport layer (e.g. \ce{TiO2}, \ce{SnO}) and hole transport layer (e.g. spiro-OMeTAD, PEDOT-PSS).
At the interface there are two key considerations.
One is that the bands should be electronically matched so as to allow efficient charge extraction without large energy loss. 
The second is that the formation of defects should be minimised as these acts as sites for recombination, can lead to mechanical degradation of the device, and have been linked to hysteresis.\cite{Almora2016a}

The commonly used hole transporter spiro-OMeTAD is hygroscopic so that stability in humid air is a concern.\cite{Tai2016}
This has prompted the development of screening procedures \cite{Butler2016a, Murray2015a} to identify alternative contacts. 
The electronic-lattice-site (ELS) figure of merit considers band alignment, lattice match and chemical viability via the overlap of atomic positions.\cite{Butler2016a}
Using this figure of merit \ce{Cu2O} is identified as a possible earth abundant hole extractor, whilst oxide perovskites such as \ce{SrTiO3} and \ce{NaNbO3} have been identified as possible electron extractors.
As with the majority of screening techniques, the candidate materials meet the necessary but perhaps not sufficient conditions. 
Further refinements may consider the change in electronic properties as lattice strain and chemical inhomogeneity at the interface is introduced.

%

\begin{table*}
\caption{\label{tab:techsol} A collection of common issues that can arise in the simulation of hybrid perovskites.
Note that for convergence of supercell size, unusual behaviour can be observed due to the fact that octahedral titling modes of perovskites are allowed in even cell expansions (e.g. $2\times2\times2$) and suppressed in odd cell expansions (e.g. $3\times3\times3$) of the cubic lattice.
Lattice dynamics is particularly sensitive to basis set convergence and plane-wave codes may require an energy cut-off 25\% higher than a typical electronic structure calculation.
For a cubic halide perovskite, $k$-point sampling of at least $6\times6\times6$ is required to give reasonable total energy and electronic structure, so a $\Gamma$-point approximation is only valid for very large supercells and should be tested carefully for the property of interest.
}
\begin{ruledtabular}
\begin{tabular}{p{5cm} p{5cm} p{7cm}}
Technique & Symptom & Solution \\
\hline
Crystal structure optimisation & Partial occupancy in structure files & Test different configurations, check total energy, and assess statistics \\
Crystal structure optimisation & Missing H in structure files & Include H based on chemical knowledge and electron counting  \\
Crystal structure optimisation & Slow ionic convergence & Try changing algorithm type and settings (rotations are poorly described by most local optimisers) \\
Electronic structure & Bandgap is too large & Include spin-orbit coupling and consider excitonic effects \\
Electronic structure & Bandgap is too small & Use a more sophisticated exchange-correlation functional \\
Electronic structure & Bandgap is still too small & Try breaking symmetry, especially for cubic perovskites \\
Electronic structure & Workfunction is positive & Align to external vacuum level using a non-polar surface  \\
Ab initio thermodynamics & No stable chemical potential range & No easy fix as many hybrid materials are metastable \\
Berry phase polarisation & Spontaneous polarisation is too large & Use appropriate reference structure and distortion pathway \\
Point defects & Negative formation energies & Check for balanced chemical reaction and chemical potential limits \\
Point defects & Transition levels are deep in bandgap & Check supercell expansion, charged defect corrections, and exchange-correlation functional  \\
Alloyed systems & Many possible configurations & Use appropriate statistical mechanics or special quasi-random structure \\
Lattice dynamics & Many imaginary phonon modes & Check supercell size and force convergence \\
Lattice dynamics & Imaginary phonon modes at zone boundaries & Use mode-following to map out potential energy surface \\
Molecular dynamics & System melts or decomposes & Check $k$-point and basis set convergence \\
Molecular dynamics & Unphysical dynamics & Check equilibration and supercell expansion \\
Molecular dynamics & No tilting observed & Use an even supercell expansion (for commensurate zone boundary phonons)  \\
Molecular dynamics & Unphysical molecular rotation rate & Check fictitious hydrogen with large mass was not used \\
Electron-phonon coupling & Values far from experiment & Consider anharmonic terms beyond linear response theory   \\
Drift-diffusion model & Current-voltage behaviour incorrect & Consider role of fluctuating ions and electrostatic potentials \\
\end{tabular}
\end{ruledtabular}
\end{table*}

\section{Conclusion}

We have outlined many of the physical properties that make hybrid perovskites unique semiconductors, but also challenging for contemporary theory and simulation. 
A number of practical points relating to issues we have encountered whilst running simulations of these materials are summarized in Table \ref{tab:techsol}.

The volume of work in this field has not allowed us to address all active areas of research, including that around perovskite-like structures with lower dimensionality (e.g. Ruddleston-Popper phases)\cite{Tsai2016,Saparov2016b,Ganose2015} and double perovskites with pairwise substitutions on the B site,\cite{Savory2016,McClure2016a,Wei2016a,Volonakis2016} 
which are both attracting significant interest. 
The optoelectronic properties of inorganic perovskites such as \ce{CsSnX3} and \ce{CsPbX3 (X = Cl, Br, I)} are also promising and provide fertile ground for future research, especially for applications in solid-state lighting.\cite{Huang2013,li2016highly}
Attempts are also being made to distil our understanding of halide perovskites into computable descriptors for large-scale screening
towards the design and discovery of novel earth-abundant non-toxic semiconductors.\cite{Brandt2015a,Davies2016,Butler2016b,Ganose2016}

There are many opportunities  ahead as we pick apart the relationship between organic and inorganic components, electronic and ionic states, as well as order and disorder in this complex family of materials. 

\acknowledgements

The primary research underpinning this discussion was performed by Jarvist M. Frost (molecular dynamic and Monte Carlo investigations), Federico Brivio (crystal and electronic structure), Jonathan M. Skelton (lattice dynamics and vibrational spectroscopy), and Lucy Whalley (bandgap deformations).
We are indebted to our large team of external collaborators including the groups of Mark van Schilfgaarde, Saiful Islam, Simon Billinge, Piers Barnes and Mark Weller.
L.W. would like to acknowledge support and guidance from the staff and students at the Centre for Doctoral Training in New and Sustainable Photovoltaics.
This work was funded by the EPSRC (grant numbers EP/L01551X/1 and EP/K016288/1), the Royal Society, and the ERC (grant no. 277757). 


%

\end{document}